\documentclass[aip, floatfix, reprint, longbibliography]{revtex4-1}
\usepackage[utf8]{inputenc}
\usepackage[T1]{fontenc}
\usepackage{enumitem}
\usepackage{amsmath}    
\usepackage{amssymb}    
\usepackage{amsthm}
\usepackage[]{algorithm}
\usepackage{algpseudocode}
\usepackage{enumitem}
\usepackage{graphicx}   
\usepackage{subfigure}  
\usepackage{hyperref}   
\usepackage[capitalise]{cleveref}   
\usepackage{tabularx}
\usepackage{booktabs}       
\usepackage{appendix}
\usepackage{bm}
\usepackage{color, soul}
\usepackage{setspace}

\newcommand{\fu}{Department of Mathematics and Computer Science, Freie Universit{\"a}t, Berlin, Germany}
\newcommand{\mpi}{IMPRS-BAC, Max Planck Institute for Molecular Genetics, Berlin, Germany}
\newcommand{\fuphys}{Department of Physics, Freie Universit{\"a}t, Berlin, Germany}
\newcommand{\stanford}{Department of Chemistry, Stanford University, Stanford, CA, USA}
\newcommand{\ricechem}{Department of Chemistry, Rice University, Houston, TX, USA} 
\newcommand{\ricebiophys}{Center for Theoretical Biological Physics, Rice University, Houston, TX, USA}
\newcommand{\ricephys}{Department of Physics, Rice University, Houston, TX, USA}
\newcommand{\barcs}{Computational Science Laboratory, Universitat Pompeu Fabra, PRBB, C/Dr Aiguader 88,  Barcelona, Spain} 
\newcommand{\barcom}{Institucio Catalana de Recerca i Estudis Avanats (ICREA), Passeig Lluis Companys 23, Barcelona, Spain} 

\begin{document}

\title{Coarse Graining Molecular Dynamics with Graph Neural Networks}

\author{Brooke E. Husic}
\email{bhusic@fu-berlin.de}
\thanks{B.E.H., N.E.C.,~and D.L.~contributed equally to this work.}
\affiliation{\fu}
\affiliation{\stanford}

\author{Nicholas E. Charron}
\thanks{B.E.H., N.E.C.,~and D.L.~contributed equally to this work.}
\affiliation{\ricephys}
\affiliation{\ricebiophys}

\author{Dominik Lemm}
\thanks{B.E.H., N.E.C.,~and D.L.~contributed equally to this work.}
\affiliation{\barcs}

\author{Jiang Wang}
\affiliation{\ricebiophys}
\affiliation{\ricechem}

\author{Adri{\`a} P{\'e}rez}
\affiliation{\barcs}

\author{Maciej Majewski}
\affiliation{\barcs}

\author{Andreas Kr{\"a}mer}
\affiliation{\fu}

\author{Yaoyi Chen}
\affiliation{\fu}
\affiliation{\mpi}

\author{Simon Olsson}
\affiliation{\fu}

\author{Gianni de Fabritiis}
\email{gianni.defabritiis@upf.edu}
\affiliation{\barcs}
\affiliation{\barcom}

\author{Frank No{\'e}}
\email{frank.noe@fu-berlin.de}
\affiliation{\fu}
\affiliation{\ricebiophys}
\affiliation{\ricechem}
\affiliation{\fuphys}

\author{Cecilia Clementi}
\email{cecilia.clementi@fu-berlin.de}
\affiliation{\fuphys}
\affiliation{\ricephys}
\affiliation{\ricebiophys}
\affiliation{\ricechem}

\begin{abstract}
Coarse graining enables the investigation of molecular dynamics for larger systems and at longer timescales than is possible at atomic resolution.
However, a coarse graining model must be formulated such that the conclusions we draw from it are consistent with the conclusions we would draw from a model at a finer level of detail.
It has been proved that a force matching scheme defines a thermodynamically consistent coarse-grained model for an atomistic system in the variational limit.
Wang \textit{et al.} [ACS Cent.~Sci.~\textbf{5}, 755 (2019)] demonstrated that the existence of such a variational limit enables the use of a supervised machine learning framework to generate a coarse-grained force field, which can then be used for simulation in the coarse-grained space.
Their framework, however, requires the manual input of molecular features upon which to machine learn the force field.
In the present contribution, we build upon the advance of Wang \textit{et al.} and introduce a hybrid architecture for the machine learning of coarse-grained force fields that learns their own features via a subnetwork that leverages continuous filter convolutions on a graph neural network architecture.
We demonstrate that this framework succeeds at reproducing the thermodynamics for small biomolecular systems.
Since the learned molecular representations are inherently transferable, the architecture presented here sets the stage for the development of machine-learned, coarse-grained force fields that are transferable across molecular systems.
\end{abstract}

\maketitle


\section{Introduction}

Technologies facilitating molecular dynamics (MD) simulations, such as distributed computing~\cite{shirts2000screen, allen2001blue, buch2010high} and bespoke hardware~\cite{shaw2008anton}, have made great strides in terms of the time- and length-scales accessible \textit{in silico}.
However, even the longest protein simulations still fail to reach total times exceeding milliseconds, and dedicated analysis methods are required to infer dynamics at longer timescales~\cite{lane2013milliseconds, plattner2017complete}.
In the context of such limitations at full atomistic resolution, coarse graining provides a crucial methodology to more efficiently simulate and analyze biomolecular systems.
In addition to the practical advantages that arise from more efficient sampling, coarse graining can also elucidate the physical components that play key roles in molecular processes.

Coarse graining is especially useful for analyzing structures and processes that reach beyond the length- and time scales accessible to all-atom MD. Important examples include protein folding, protein structure prediction, and protein interactions~\cite{kmiecik2016coarse}. Some of the most-used coarse-grained models for such studies are structure-based models~\cite{clementi2000topological}, MARTINI~\cite{marrink2007martini, monticelli2008martini}, CABS~\cite{kolinski2004protein}, AWSEM~\cite{davtyan2012awsem}, and Rosetta~\cite{das2008macromolecular}.
These models differ with respect to their potential energy function, parameterization approaches, and resolution, which in combination determine their efficiency, accuracy, and transferability.
In the past decade, coarse-grained models have become increasingly powerful due to an unprecedented wealth of experimental reference data and computational capabilities.
In this context, the development of more realistic architectures and modeling approaches is of prime importance.

In the field of computer science, advances in hardware and autodifferentiation software have enabled enormous progress in machine learning algorithms, including at the intersection of computation and the physical sciences~\cite{noe2020machine, gomez2020machine}.
Crucial to the use of neural networks in the physical sciences is a consideration for the form the training data takes before it is input into the network.
One strategy for representing molecules mathematically is through the use of graphs, whose nodes and edges intuitively correspond to atoms and bonds of (or interatomic distances within) a molecule, respectively.
By performing multiple convolution operations on a graph, each node can influence other increasingly distant nodes.
The use of \textit{graph neural networks}~\cite{kipf2017semi, battaglia2018relational} in the molecular sciences is therefore a promising direction in a variety of applications, and graph convolutional architectures have been used to predict molecular~\cite{duvenaud2015convolutional, kearnes2016molecular, gilmer2017neural, feinberg2018potentialnet} and material~\cite{xie2018crystal} properties as well as atomic energies~\cite{schutt2017quantum, schutt2017schnet} and forces~\cite{ruza2020temperature}.

In this work, we combine the use of graph representations of molecules with a supervised neural network architecture in the coarse graining context.
We consider coarse graining to be the process of reducing structural degrees of freedom to facilitate more efficient simulation with specific goals in mind (e.g., reproducing system thermodynamics).
Coarse graining can be implemented with a ``top down'' or ``bottom up'' approach, although other categories can be determined and strategies can be combined~\cite{noid2013perspective}.
In a ``top down'' scheme, coarse graining frameworks are explicitly designed to reproduce certain macroscale emergent properties~\cite{noid2013perspective}.
In a ``bottom up'' framework, which we consider here, implementations focus instead on reproducing specific features from a more detailed model.

The latter involves (i)~a mapping from the entities in a fine-grained (e.g., atomistic) representation to a smaller set of interaction sites, often called ``beads,'' and (ii)~a physical model (i.e., Hamiltonian function) for the coarse-grained system comprising those beads.
Good choices for the mapping and model will lead to more efficient simulation while preserving the biophysical properties of interest to the researcher.
Modern machine learning techniques have been recently employed to learn both the mapping~\cite{boninsegna2018data, wang2019coarse} and the model~\cite{john2017many, zhang2018deepcg, wang2019machine, wang2020ensemble, ruza2020temperature} components of bottom up coarse graining.

In the present contribution, we focus on the coarse graining model and employ a bottom up ``force matching'' scheme formulated as a supervised machine learning problem to reproduce the thermodynamics of small biomolecular systems.
Particularly, we modify the architecture of the recently-introduced CGnet framework~\cite{wang2019machine} such that the molecular features it requires are \emph{learned} via graph convolutional neural networks instead of hand-selected as in the original formulation.
By leveraging the inherently transferable SchNet scheme~\cite{schutt2017schnet, schutt2018schnet} to learn features, we render the entire CGnet framework transferable across molecular systems.

Our goal in this paper is to present the theory underlying CGSchNet---our new transferable coarse graining architecture---and to demonstrate its success on learning the thermodynamics of individual biomolecular systems.
We find that our new protocol produces more accurate free energy surfaces in comparison with the use of hand-selected features, is more robust to hyperparameter choices, and requires less regularization.
Presented alongside a machine learning software package that implements the methods introduced, the current contribution sets out a framework for the machine learning of transferable, coarse-grained molecular force fields and demonstrates its application to a small peptide system and the miniprotein chignolin~\cite{honda2008crystal}.
The practical application of the methods described herein to larger protein systems, particularly those characterized by meaningful tertiary structure, remains an open challenge that will be explored in future work.


\section{Theory} \label{sec:theory}

Force matching was pioneered in the atomistic setting, in which forces obtained from an inexpensive calculation are matched to forces computed at a more computationally expensive level of theory (i.e., quantum) via an optimization scheme~\cite{ercolessi1994interatomic}.
The method was later adapted by the coarse graining community; in that context, coarse-grained representations are sought such that the forces computed from the coarse-grained energy function for a given configuration match the average forces on corresponding atomistic representations~\cite{izvekov2005multiscale}.

Because coarse graining away degrees of freedom entails that multiple atomistic structures will correspond to the same coarse-grained configuration, it is impossible to obtain zero error during force matching in the coarse graining context.
However, it can be proved that the coarse graining model that matches the \emph{mean} forces yields the correct thermodynamics, and that the objective is variationally bounded from below by a value that necessarily exceeds zero.

In Sec.~\ref{sec:theory}, we overview the major advances that enable the present contribution.
The practically inclined reader may proceed directly to Sec.~\ref{sec:methods}, where we discuss the CGnet architecture and introduce this work's methodological contribution:~namely, the incorporation of learnable molecular features into CGnet via the use of continuous filter convolutions on a graph neural network (i.e., SchNet~\cite{schutt2017schnet, schutt2018schnet}).
We will see in Sec.~\ref{sec:methods} that the scheme we introduce here enables, at least in principle, a coarse graining architecture that is transferable across system size and sequence.
The practical use of this architecture to learn a force field in a transferable context will be addressed in future work.

\subsection{Force matching} \label{subsec:force_matching}

Consider an all-atom dataset of coordinates and corresponding forces, which we have obtained using a high level calculation (e.g., \textit{ab initio}).
We denote each three-dimensional structure $\mathbf{r}_i \in \mathbb{R}^{3N}$, $i = 1, \dots, M$, and the forces $\mathbf{F}(\mathbf{r}_i) \in \mathbb{R}^{3N}$, where $N$ is the number of atoms in the system.
Now consider a trial energy function $\hat{V}(\mathbf{r}_i; \mathbf{\Theta})$, which takes as arguments an atomistic configuration $\mathbf{r}_i$ and any parameters $\mathbf{\Theta}$.
We would like to use $\hat{V}$ to predict the forces on every $\mathbf{r}_i$---presumably in a more efficient way---by taking its negative derivative.
We can write the ``force matching'' problem of comparing the two sets of atomistic forces as,
\begin{align}
    L(\mathbf{R}; \mathbf{\Theta}) &= \frac{1}{3MN} \sum_{i=1}^M \Vert \underbrace{\mathbf{F}(\mathbf{r}_i)}_{\substack{\text{``True''} \\ \text{forces}}} + \underbrace{\nabla_{\mathbf{r}_i} \hat{V} (\mathbf{r}_i; \mathbf{\Theta})}_{\substack{\text{(Negative)} \\ \text{predicted}\\ \text{forces}}} \Vert ^2, \label{eq:force_matching_loss}
\end{align}

\noindent{}where $\mathbf{R}$ is the set of all $M$ sampled atomistic configurations.

The objective~\eqref{eq:force_matching_loss} was introduced by Ercolessi and Adams to analyze \textit{ab initio} simulations of elemental aluminum~\cite{ercolessi1994interatomic}.
The authors highlight the method's need to accommodate invariant properties of the system and discuss the requirement of a variety of geometries, physical conditions, and system identities in $\mathbf{R}$ if the learned potential is to be transferable across conformation, thermodynamic, or chemical space, respectively.
Subsequent work has derived analytical approaches to this scheme in the context of liquids~\cite{izvekov2004effective, guenza2018accuracy}.

A decade later, Izvekov and Voth introduced the multiscale coarse graining (MS--CG) method, a groundbreaking advance that adapts force matching to the coarse graining context~\cite{izvekov2005multiscale, izvekov2005liquid}.
The MS--CG framework involves two steps:~first, atoms are aggregated into ``interaction sites'' according to a linear mapping from $N$ atoms to $n$ interaction sites (henceforth ``beads''),

\begin{align}
    \mathbf{x}_i = \mathbf{\Xi}\mathbf{r}_i \in \mathbb{R}^{3n}, \label{eq:cg_map}
\end{align}

\noindent{}where $\mathbf{x}_i$ is the coarse-grained representation with $n < N$ beads and the matrix $\mathbf{\Xi} \in \mathbb{R}^{3n \times 3N}$ effectively performs a clustering from the original atoms to the beads.
Then, force matching is performed between a transformation of the atomistic forces and a set of predicted coarse-grained forces.
This procedure thereby creates a ``multiscale'' link between the all-atom and coarse-grained representations~\cite{izvekov2005multiscale}.

Consider a \emph{coarse-grained} energy function $U(\mathbf{x}; \mathbf{\Theta})$.
Let's say we have a set of $M$ coarse-grained configurations that we have obtained by applying~\eqref{eq:cg_map} to every configuration $\mathbf{r}_i \in \mathbf{R}$.
To calculate the forces on the beads, we then take the negative derivative of $U$ with respect to the reduced coordinates; in other words, we evaluate,

\begin{align*}
-\nabla U (\mathbf{\Xi}\mathbf{r}_i; \mathbf{\Theta}) =
-\nabla_{\mathbf{x}_i} U (\mathbf{x}_i; \mathbf{\Theta})
\in \mathbb{R}^{3n},
\end{align*}
\noindent{}for each configuration $i$.
From here we have all the ingredients to write down the adaptation of~\eqref{eq:force_matching_loss} to the MS--CG method:

\begin{align}
    L(\mathbf{R}; \mathbf{\Theta}) &= \frac{1}{3Mn} \sum_{i=1}^M \Vert \underbrace{\mathbf{\Xi_F}\mathbf{F}(\mathbf{r}_i)}_{\substack{\text{Atomistic forces} \\ \text{mapped to} \\ \text{coarse-grained} \\ \text{space}}} + \underbrace{\nabla U (\mathbf{\Xi}\mathbf{r}_i; \mathbf{\Theta})}_{\substack{\text{(Negative) forces} \\ \text{predicted from }\\ \text{coarse-grained} \\ \text{model} }} \Vert ^2. \label{eq:mscg_loss}
\end{align}

\noindent{}where $\mathbf{\Xi_F}\mathbf{F}$ is the \textit{instantaneous} coarse-grained force (also called the \textit{local mean force}); that is, the projection of the atomistic force into the coarse-grained space. A general expression for the force projection~\cite{ciccotti2008projection} is $\mathbf{\Xi_F} = (\mathbf{\Xi}\mathbf{\Xi}^\top)^{-1} \mathbf{\Xi}$. Other choices for the mapping $\mathbf{\Xi_F}$ are possible and used for coarse graining~\cite{noid2008multiscale}.

In principle, the coarse-grained energy $U(\mathbf{x})$ that is \textit{exactly} thermodynamically consistent with the atomistic energy $V(\mathbf{r})$ can be expressed analytically as:

\begin{align}
    U(\mathbf{x}) &= -k_B T \ln p^{\text{CG}} (\mathbf{x}) + \text{Constant}, \label{eq:cg_energy}
\end{align}

\noindent{}where $k_B$ is Boltzmann's constant and $T$ is the absolute temperature. The function $p^{\text{CG}}$ is the marginal probability density,

\begin{align}
    p^{\text{CG}}(\mathbf{x}') = \frac{\int_{\mathcal{R}} \exp\left(-\frac{V(\mathbf{r})}{k_B T}\right)\delta(\mathbf{x}' - \mathbf{\Xi}\mathbf{r})d\mathbf{r}}{\int_{\mathcal{R}} \exp \left(-\frac{V(\mathbf{r})}{k_B T}\right) d\mathbf{r}}, \label{eq:analytical_distr}
\end{align}

\noindent{}where $\mathcal{R}$ is the set of all possible atomistic configurations.
Since we are concerned with all theoretically accessible structures and (thus) employ an integral formulation, we have dropped the subscripts $i$ with the understanding that $\mathbf{x}'$ and $\mathbf{r}$ now refer to infinitesimally small regions of their respective configuration spaces.
$\mathbf{x}'$ is distinguished from $\mathbf{x}$ to emphasize that~\eqref{eq:analytical_distr} is substituted into~\eqref{eq:cg_energy} as a function, not a number.

The coarse-grained energy function~\eqref{eq:cg_energy} is called the potential of mean force (PMF) and is an analogue of the atomistic potential energy function.
Via~\eqref{eq:analytical_distr}, it is a function of weighted averages of energies of atomistic configurations.
For a given coarse-grained structure $\mathbf{x}'$, in~\eqref{eq:analytical_distr} we evaluate whether every possible $\mathbf{r} \in \mathcal{R}$ maps to $\mathbf{x}'$.
We expect multiple atomistic configurations $\mathbf{r}$ to map to $\mathbf{x'}$ due to the reduction in degrees of freedom that results from structural coarse graining (n.b., this means the PMF is in fact a free energy, as it contains entropic information~\cite{noid2013perspective}).
In these cases, the Dirac delta function in~\eqref{eq:analytical_distr} returns one, and the contribution of that atomistic configuration to the marginal probability distribution is a function of its Boltzmann factor.
If $\mathbf{r}$ does not map to $\mathbf{x}'$, then the evaluation of the delta function (and thus the contribution of that atomistic structure to the free energy of $\mathbf{x}'$) is zero.
The denominator of the right-hand side of~\eqref{eq:analytical_distr} is the all-atom partition function, which serves as a normalization factor.

To calculate the forces on our coarse-grained beads, we must take the gradient of~\eqref{eq:cg_energy}.
However, since we cannot exhaustively sample $\mathcal{R}$,~\eqref{eq:analytical_distr} is intractable, and we must approximate $U$ instead.
One way to approximate $U$ is to employ force matching---that is, by minimizing~\ref{eq:mscg_loss}---as we describe in Sec.~\ref{subsec:cgnet}.
Another method, which we do not discuss in this report, is through relative entropy~\cite{shell2008relative}, whose objective is related to that of force matching~\cite{rudzinski2011coarse, noid2013perspective}.

\subsection{Coarse graining as a supervised machine learning problem} \label{subsec:cgnet}

In 2008, \citet{noid2008multiscale} formalized the notion of thermodynamic consistency and established the conditions under which it is guaranteed by the MS--CG approach:~namely, that thermodynamic consistency is achieved when the coarse-grained coordinates are a linear combination of the all-atom coordinates (cf.~\eqref{eq:cg_map}) and that the equilibrium distribution of the coarse-grained configurations is equal to the one implied by the equilibrium distribution of the atomic configurations (cf.~\eqref{eq:cg_energy}).
Noid \textit{et al.}~then prove that, under certain restrictions of the coarse-grained mapping, the coarse-grained potential that achieves thermodynamic consistency at a given temperature is unique (up to an additive constant, cf.~\eqref{eq:analytical_distr})~\cite{noid2008multiscale}.

The authors define an error functional that is (uniquely) minimized for the thermodynamically consistent coarse-grained force field~\cite{noid2008multiscale, noid2008multiscale2}.
This framework provides the variational principle underlying the MS--CG method.
It follows that a variational approach can be used to search for the consistent coarse-grained force field.\footnote{
In practice, such a search is limited by the basis of trial force fields as well as the finite simulation data used~\cite{noid2008multiscale2}.
}
The variational principle entails that we can refer to~\eqref{eq:force_matching_loss} and~\eqref{eq:mscg_loss} as ``loss functions'' because they return a scalar that assumes a minimum value on the optimal model.
In recent reports from both Wang, Clementi, \textit{et al.}~\cite{wang2019machine} and \citet{wang2019coarse}, this fact is leveraged to formulate coarse graining via force matching as a supervised or semi-supervised machine learning problem, respectively.
Here, we build upon on the supervised learning case introduced in Ref.~\onlinecite{wang2019machine} as \textit{CGnet}.

In their study, Wang \textit{et al.}~present several crucial contributions~\cite{wang2019machine}.
First, they decompose the error term implied by~\eqref{eq:mscg_loss} into three physically meaningful components; namely, bias, variance, and noise.
Second, the authors introduce \emph{CGnet}:~a neural network architecture designed to minimize the loss in~\eqref{eq:mscg_loss}.
Once a CGnet is trained, it can be used as a force field for new data points in the coarse-grained space  while enforcing known properties of the system such as symmetries and equivariances (see Sec.~\ref{subsec:simulation}).
Third, Wang \textit{et al.} augment their initial framework to introduce regularized CGnets~\cite{wang2019machine}.
Regularized CGnets avoid catastrophically wrong predictions observed in their ``unregularized'' counterparts by introducing the calculation of prior energy terms before training.
This adjustment means that, instead of learning the forces directly, the neural network learns a \emph{correction} to the prior terms in order to match the atomistic forces.

Using regularized CGnets (henceforth, we assume all CGnets are regularized) on two peptide systems, the authors demonstrated effective learning of coarse-grained force fields that could not be obtained with a few-body model approach~\cite{wang2019machine}.
It is from this baseline that we present CGSchNet, an augmentation of the CGnet methodology.

\section{Methods} \label{sec:methods}

In the quantum community, supervised machine learning has been used to predict energies on small molecules through a variety of approaches~\cite{behler2007generalized, bartok2010gaussian, rupp2012fast, bartok2013machine, smith2017ani, chmiela2017machine, bartok2017machine, schutt2017quantum, smith2018less, schutt2018schnet, grisafi2018symmetry, imbalzano2018automatic, nguyen2018comparison, zhang2018end, zhang2018deep, bereau2018non, wang2018toward}.
In particular, the SchNet architecture is based on the use of continuous filter convolutions and a graph neutral network~\cite{battaglia2018relational, schutt2017schnet, schutt2018schnet}.
SchNet is a scalable, transferable framework that employs representation learning to predict the properties and behavior of small organic molecules.
In the vein of the original force matching procedure of \citet{ercolessi1994interatomic}, SchNet has also been used to predict forces on atomic data from a quantum mechanical gold standard~\cite{schutt2018schnet}.

In Sec.~\ref{subsec:cgnet_architecture} we briefly overview the CGnet scheme upon which we base the method introduced in this work.
Then, in Sec.~\ref{subsec:schnet_features}, we describe SchNet and introduce our adaptation of SchNet to the coarse graining problem by incorporating it into a CGnet to create a hybrid ``CGSchNet'' architecture.
The original implementation of CGnet is not transferable across different systems due to its reliance on hand-selected structural features~\cite{wang2019machine}.
We recognized that SchNet could be leveraged as a subcomponent of CGnet in order to \emph{learn} the features, thereby converting CGnet---i.e., force matching via supervised machine learning---to a transferable framework for the first time. 

\subsection{Original CGnet architecture} \label{subsec:cgnet_architecture}

For both CGnet and CGSchNet, our training data comprises an MD simulation that has already been performed and for which the atomistic forces have been retained or recomputed.
Both the configurations and the forces are in $\mathbb{R}^{3N}$ space for $N$ atoms.
We then determine our mapping matrix $\mathbf{\Xi}$ and use it to prepare our input data (coarse-grained structures) and labels (atomistic forces mapped to the coarse-grained space), which will both be in $\mathbb{R}^{3n}$ for $n$ beads (recall~\eqref{eq:cg_map}).

While the mapping is permitted to be more general, in our work we restrict it to the special case where the matrix $\mathbf{\Xi}$ contains zeroes and ones only.
With this choice of mapping, the projection of the forces in~\eqref{eq:mscg_loss} becomes simply $\mathbf{\Xi}_\mathbf{F} = \mathbf{\Xi}$.
Our mapping thus ``slices'' the original atomic configuration such that the corresponding coarse-grained representation comprises a subset of the original atoms.
For example, a useful mapping might retain only protein backbone atoms or $\alpha$-carbons.

To construct a CGnet, the structural data is preprocessed such that it is represented by features with the desired properties.
Wang \textit{et al.}~use a set of distances, planar angles, and torsional angles~\cite{wang2019machine}. 
In the present work, on the other hand, instead of using hand-selected structural features, we require only distances and bead identities from which features are \emph{learned}; this is described in Sec.~\ref{subsec:schnet_features}.

For their regularized implementation, Wang \textit{et al.} use up to two types of prior terms in CGnets~\cite{wang2019machine}.
The first is a harmonic prior on selected distances (i.e., bonds or pseudobonds) and angles.
The second is a repulsion prior that can be used on nonbonded distances.
Respectively, these priors are defined as follows for a given feature $f_i$ calculated from the data (e.g., a particular distance),

\begin{subequations}
\begin{align}
    U_i^\text{harmonic}(f_i) &= \frac{k_B T}{2 \text{Var}[f_i]} \Large(f_i - \mathbb{E}[f_i]\Large)^2, \\
    U_i^\text{repulsion}(f_i) &= \left(\frac{\sigma}{f_i}\right)^c. \label{eq:repul_prior}
\end{align}
\end{subequations}

\noindent{}The constants in~\eqref{eq:repul_prior} can be determined through cross-validated hyperparameter optimization as in Ref.~\onlinecite{wang2019machine}.
The prior energy is the sum of each prior term for all relevant features  $f_i$.
In principle, any scalar function of protein coordinates can be used to construct a prior energy term.

The original CGnet uses a fully connected network to learn corrections to the prior energy~\cite{wang2019machine}.
Crucially, the last layer of the network returns a scalar output.
Because of this single node bottleneck structure, the resulting coarse-grained force field will be curl-free and is therefore guaranteed to conserve energy~\cite{chmiela2017machine, wang2019machine}.
Since all the steps described are differentiable, we can use an autodifferentiation framework such as PyTorch~\cite{paszke2019pytorch} to take the derivative of the energy with respect to the original (coarse-grained) spatial coordinates via backpropagation.
This derivative corresponds to the predicted forces on the coarse-grained beads in $\mathbb{R}^{3n}$, which can then be compared to the known forces on the training coordinates.

\subsection{Replacing structural features with graph neural networks} \label{subsec:schnet_features}

Wang \textit{et al.}~show that CGnets constructed upon hand-selected structural features produce machine-learned force fields yielding accurate free energy surfaces~\cite{wang2019machine}.
The model architecture is found to be somewhat sensitive to various hyperparameters and required individual tuning for each system (see e.g.~Fig.~5 in \citet{wang2019machine}).
Furthermore, a new system will in general require retraining because the feature size is fixed according to the system geometry.

In the present contribution, we replace the fixed structural features employed in the original CGnet formulation (i.e., distances, angles, and torsions)~\cite{wang2019machine} with \emph{learned} features computed using continuous filter convolutions on a graph neural network (SchNet~\cite{schutt2017schnet, schutt2018schnet}).
The SchNet architecture thereby becomes a subunit of CGnet with its own, separate neural network scheme; we refer to this hybrid architecture as \emph{CGSchNet}.

The term graph neural network was introduced in \citet{battaglia2018relational} as a generalization for networks operating on graph structures, including but not limited to graph convolutional networks~\cite{kipf2017semi} and message passing
networks~\cite{gilmer2017neural}.
These networks have in common that they have a notion of $n$ nodes $\mathcal{V}$ connected by edges $\mathcal{E}$ in a graph structure.
In each neural network layer, information is passed between nodes and representations of the nodes and/or edges are updated.
The various types of graph neural networks differ according to whether there are node updates, edge updates, or both; as well as by how functions are shared across the network and how the network output is generated.
A fairly general formulation of a graph neural network with node updates is as follows:~each node
$i$ is associated with an initial node representation $\mathbf{h}^{(0)}_i$; in other words, $\mathbf{h}^{(0)}_i$ is a vector that represents the type or identity of the node. In each layer of the neural network, the node representations are updated according to,
\begin{equation}
\label{eq:gnn_node_update}
    \mathbf{h}^{(t+1)}_i = f[(\mathbf{h}^{(t)}_j), (\mathbf{e}_{ij})],
\end{equation}
where $i, j = 1, \dots, n$, $\mathbf{e}_{ij}$ are edge features  defined for all edges in $\mathcal{E}$, and $f$ is a  trainable neural network function.
After $T$ such layers, an output is generated,
\begin{equation}
\label{eq:gnn_output}
    \mathbf{o} = g[(\mathbf{h}^{(T)}_j), (\mathbf{e}_{ij})].
\end{equation}
In the present contribution, we make the following choices:

\begin{enumerate}
    \item Graph nodes represent coarse-grained beads.
    \item Because multi-body interactions are important for the coarse graining problem, edges are defined between all beads (or all beads within a specified cutoff radius).
    \item The edge features $\mathbf{e}_{ij}$ are taken to be the distances between beads, implying translational and rotational invariance of the network output.
    \item The update function $f$ in~\eqref{eq:gnn_node_update} is chosen to be a continuous convolution update as in SchNet~\cite{schutt2017schnet}.
    \item The entire trainable part of a CGnet~\cite{wang2019machine}---in this case a beadwise multilayer perceptron/dense neural network---becomes the output function $g$ in ~\eqref{eq:gnn_output}. Because this output is beadwise the learnable coarse grain energy is invariant with respect to permutations of identical beads.
    \item The output $\mathbf{o}$ is a scalar; namely, the coarse-grained energy before the addition of the prior energy term. 
\end{enumerate}

Below we describe the SchNet updates in more detail and
how to incorporate SchNet into CGnet to create CGSchNet.

\subsubsection{Learning molecular representations with SchNet} \label{subsec:schnet}

One key motivating factor for the original development of SchNet is that, unlike the images and videos that comprise the datasets for much of modern machine learning, molecular structures are not restricted to a regular grid.
Therefore, Sch{\"u}tt \textit{et al.}~introduced continuous-filter convolutions to analyze the structures of small molecules with the goal of predicting energies and forces according to a quantum mechanical gold standard~\cite{schutt2017schnet}.
This development builds upon previous work in predicting atomic properties directly from structural coordinates~\cite{chmiela2017machine, schutt2017quantum, gilmer2017neural}.

SchNet is a graph neural network where the nodes correspond to particles embedded in three-dimensional space and the convolutional filters depend on interparticle distances, which preserves invariances expected in the system~\cite{schutt2017schnet, battaglia2018relational}. 
While SchNet was originally used to predict quantum-chemical energies from atomistic representations of small molecules, here we employ it to learn a feature representation that replaces the hand-selected features in a CGnet for the purpose of predicting the coarse-grained energy on the coarse-grained bead coordinates $\mathbf{x}_i$.

\begin{figure}[t!]
\centering
\includegraphics[width=0.5\textwidth]{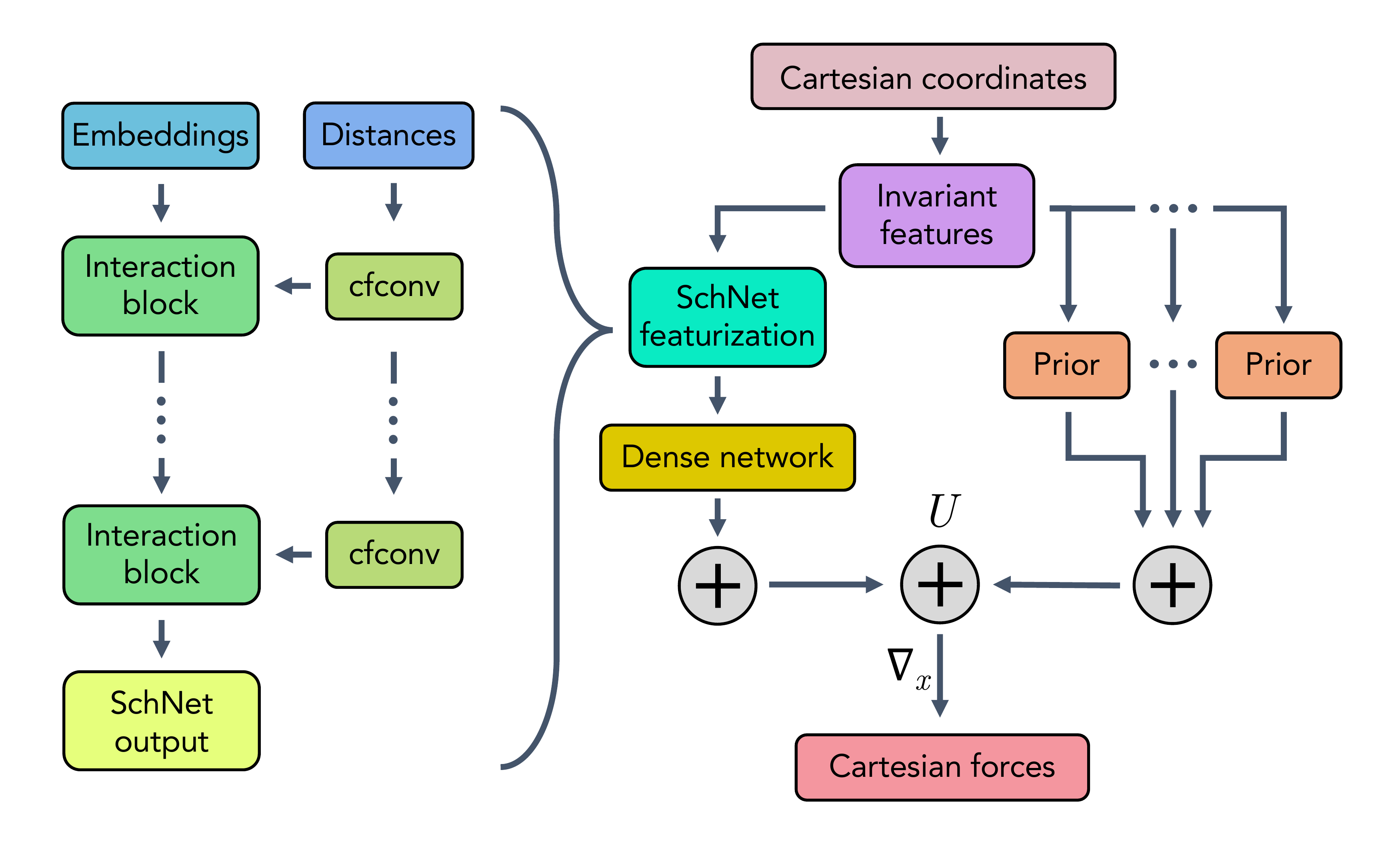}
\caption{
CGSchNet architecture.
}
\label{fig:architecture}
\end{figure}

As in other graph neural networks, SchNet learns feature vectors on the nodes (here, coarse-grained beads).
The initial node features at the input are called node or bead embeddings $\mathbf{h}_i^{(0)}$, which are given by trainable, shared vectors with $d_h$ dimensions (``Embeddings'' in Fig.~\ref{fig:architecture}),
\begin{equation}
    \mathbf{h}_i^{(0)} = \mathbf{a}_{k(i)}.
\end{equation}
Here, $k(i)$ is a lookup table that maps the bead index
$i$ to its type $k$. In the present applications we  use nuclear charges (capped alanine) or amino acid identities (chignolin) as bead types. 
The bead embeddings are shared among beads of the same type and are optimized during training. 
Crucially, this entails that SchNet \emph{learns} a molecular representation, which avoids the common paradigm of fixed, heuristic feature representations.

Next, we describe how bead representations are updated (``Interaction block'' and ``cfconf'' in Fig.~\ref{fig:architecture}).
In our current architecture, these updates are implemented as in the original SchNet unless noted otherwise (see Refs.~\onlinecite{schutt2017schnet} and~\onlinecite{schutt2018schnet} for details).

In each interaction layer, we perform a continuous
convolution between beads.
For this, the inter-bead distances $| \mathbf{x}_j - \mathbf{x}_i |$ are featurized using radial basis functions $\mathbf{e}$, e.g., one-dimensional Gaussians centered at different distances.
These featurized distances serve as the input to a filter-generating neural network $w$ that maps the featurized distance input $\mathbf{e}(| \mathbf{x}_j - \mathbf{x}_i |)$ to a $d_h$-dimensional filter.
This filter is applied to the bead representations $\mathbf{h}_i$ as follows (``cfconf'' in Fig. \ref{fig:architecture}),
\begin{equation}
\label{eq:cfconf}
    \mathbf{z}_i^{(t)} = \sum_j w^{(t)}[\mathbf{e}(| \mathbf{x}_j - \mathbf{x}_i |)] \cdot b^{(t)}(\mathbf{h}_i^{t})
\end{equation}
Here, $w$ and $b$ are trainable functions and $\cdot$ is element-wise multiplication.
As in the original SchNet implementation~\cite{schutt2017schnet}, $w$ is a dense neural network and $b$ is a beadwise linear layer.
The sum in~\eqref{eq:cfconf} is taken over every bead $j$ within the neighborhood of bead $i$, which can be all other beads in the system or a subset thereof if a finite neighborhood is specified.
Even when interactions are limited to particles within a cutoff radius, a sequence of multiple interaction layers will eventually allow all particles to be interacting, and therefore be able to express complex multi-body interactions.

In each layer, the bead representations are updated in interaction blocks, each of which comprises a residual update of the bead representation via a nonlinear function of the continuous convolution outputs $\mathbf{z}_i^{(t)}$ (``interaction block'' in Fig.~\ref{fig:architecture}),
\begin{equation}
    \mathbf{h}_i^{(t+1)} = \mathbf{h}_i^{(t)} + g^{(t)}(\mathbf{z}_i^{(t)}).
\end{equation}
The residual update step is an ``additive refinement'' that prevents gradient annihilation in deep networks \cite{he2016deep}.
As described by Sch{\"u}tt \textit{et al.}~\cite{schutt2017schnet, schutt2018schnet}, the trainable function $g$ involves beadwise linear layers and a nonlinearity. Instead of the softplus nonlinearity used in the original SchNet~\cite{schutt2017schnet}, here we use the hyperbolic tangent.

Following the last interaction layer we must choose an
output function (\ref{eq:gnn_output}).
As in the original SchNet implementation, the output of the final SchNet interaction block is input into an beadwise CGnet multilayer perceptron/dense network.
An important feature of transferability is permutation invariance of beads with identical type.
In the context of coarse graining, this means the contribution of a bead to the coarse-grained energy should depend on its location in the molecular graph, but not at which index this bead is positioned in the input.
SchNet layers are permutation-equivariant, i.e., any exchange of the input representations $\mathbf{h}^{(0)}_i$ will correspond to the same exchange of the learned representations $\mathbf{h}^{(T)}_i$.
In order to obtain permutation-invariant energies (as in the original SchNet publications~\cite{schutt2017schnet, schutt2018schnet}), the beadwise output CGnet network contracts down to a scalar energy prediction that is then summed over all beads to yield the total learnable part of the coarse-grained energy. It is important to note that the models used in this study employ priors that are not permutation invariant, and so the non-learnable part of the coarse-grained energy (i.e., the prior terms) breaks permutation invariance in the model overall. The development of permutation invariant priors is left for future work.

In the present paper we do not use CGSchNet in a transferable manner, but rather demonstrate its capabilities when trained on individual molecular systems as a foundation for future work.
For this reason, here we use a (regularized) beadwise CGnet as the output function; i.e., a beadwise multilayer perceptron/dense neural network at whose output the learned part of the coarse-grained energy is predicted. In so doing, the SchNet interaction layers learn the input representation for a beadwise CGnet, and the beadwise CGnet ``fine-tunes'' the bead energies predicted by SchNet.

\subsubsection{CGSchNet: a transferable architecture for coarse graining} \label{subsec:cgschnet}

CGnet as originally presented is incapable of learning a transferable coarse-grained force field due to its reliance upon system-specific structural features~\cite{wang2019machine}.
Since SchNet is inherently a transferable framework, learning CGnet features using SchNet enables the transferability of the entire CGnet architecture across molecular systems.
Here, we present the advance of incorporating SchNet~\cite{schutt2017schnet, schutt2018schnet} into CGnet to replace hand-selected features with machine-learned ones.

In CGSchNet, instead of predetermined structural features---i.e., distances, angles, and torsions---a SchNet is used instead, enabling the model to learn the representation itself (see Fig.~\ref{fig:architecture}).
By replacing fixed-size geometric features with SchNet, we obtain a more flexible representation that both scales better with system size and is amenable to a transferable architecture~\cite{schutt2018schnet}.
While angles and torsions may still be included in the prior energy terms, they are no longer propagated through any neural networks.

The use of SchNet requires us not only to provide structural coordinates but also a type for every bead.
In the original (i.e., non-coarse graining) implementation for systems at atomic resolution, the types are atomic numbers~\cite{schutt2017quantum, schutt2017schnet, schutt2018schnet}.
In the new context presented here (i.e., leveraging SchNet for coarse graining), we may specify coarse-grained bead types---effectively, chemical environments---however we deem appropriate for the system under study; for example, amino acid identities may be used.

One can view the typing requirement of the SchNet architecture as the outlet through which to incorporate physical or chemical intuition about the system into the model, as opposed to through fixed structural features.
The benefit of the SchNet choice is that it enables an architecture that is transferable across size and sequence space because a set of embeddings can apply to multiple systems with the same components (e.g., atoms as in SchNet~\cite{schutt2017schnet} or amino acid types in proteins), whereas hand-selected structural features are not transferable across different systems. 

Finally, we note that we train CGSchNet with the coarse-grained force matching loss~\eqref{eq:mscg_loss}, which compares our predicted forces to the known forces from the training set.
Unlike in the original SchNet formulation~\cite{schutt2017schnet}, we cannot straightforwardly incorporate an additional ``energy matching'' term into the coarse graining framework.
This is because we do not have labels for the coarse-grained free energies:~these energies are defined by an integral over all microscopic configurations associated with the same coarse-grained configuration (cf.~\eqref{eq:cg_energy}), and this integral cannot be solved exactly.

\subsection{Coarse-grained simulations} \label{subsec:simulation}

A trained CGSchNet can be used as a force field to simulate the system in the coarse-grained space.
Specifically, Langevin dynamics~\cite{schneider1978molecular, fass2018quantifying} are employed to propagate coarse-grained coordinates $\mathbf{x}_t$ forward in time according to,
\begin{align} \label{eq:langevin}
    \frac{\partial^2 \mathbf{x}_t}{\partial t^2} &= -\mathbf{M}^{-1}\nabla U (\mathbf{x}_t) - \gamma \frac{\partial \mathbf{x}_t}{\partial t} + \sqrt{2k_b T \gamma}\mathbf{M}^{-\frac{1}{2}}\mathbf{W}(t),
\end{align}
\noindent{}where the diagonal matrix $\mathbf{M}$ contains the bead masses, $\gamma$ is a collision rate with units ps$^{-1}$, and $\mathbf{W}(t)$ is a stationary Gaussian process with $\langle W(t)\rangle = 0$ and $\langle W(t)W(t') \rangle = \delta(t - t')$, where $\langle \cdot \rangle$ is the mean.
In practice, we integrate~\eqref{eq:langevin} using a ``BAOAB'' Langevin integrator~\cite{leimkuhler2013rational}, and the integral of $W(t)$ is a Wiener process.

A special case of Langevin dynamics are so-called ``overdamped'' Langevin dynamics, also referred to as Brownian dynamics. 
Overdamped Langevin dynamics lack inertia. 
After setting the acceleration to zero, dividing both sides by $\gamma$, and rearranging terms, in the overdamped case,~\eqref{eq:langevin} becomes,
\begin{align} \label{eq:overdamped}
    \frac{\partial \mathbf{x}_t}{\partial t} = -\frac{\mathbf{D}}{k_B T}\nabla U(\mathbf{x}_t) + \mathbf{D}^{\frac{1}{2}}\sqrt{2 }\mathbf{W}(t),
\end{align}
\noindent{}where the diffusion matrix $\mathbf{D} \equiv \mathbf{M}^{-1} k_B T /\gamma$.
Although $\mathbf{D}$ contains a notion of mass, we note that propagating coarse-grained dynamics via~\eqref{eq:overdamped} does not actually require bead masses, since the product $\mathbf{M}\gamma$ can be considered without separating its factors.
\citet{wang2019machine} use exclusively~\eqref{eq:overdamped} with the Euler method to simulate dynamics from CGnets, with a constant diffusion matrix proportional to the identity matrix. 

In both formulations, the noise term is intended to indirectly model collisions---e.g., from and among solvent particles---that are not present in the coarse-grained coordinate space.
Since Langevin dynamics depend only on the coordinates (and, unless overdamped, velocities) of the previous time step, these simulations can easily be run in parallel from a set of initial coordinates.
The resulting coarse-grained simulation dataset can then be used for further analysis as we will show in Sec.~\ref{sec:results}.


\begin{figure}[h!]
\centering
\includegraphics[width=0.4\textwidth]{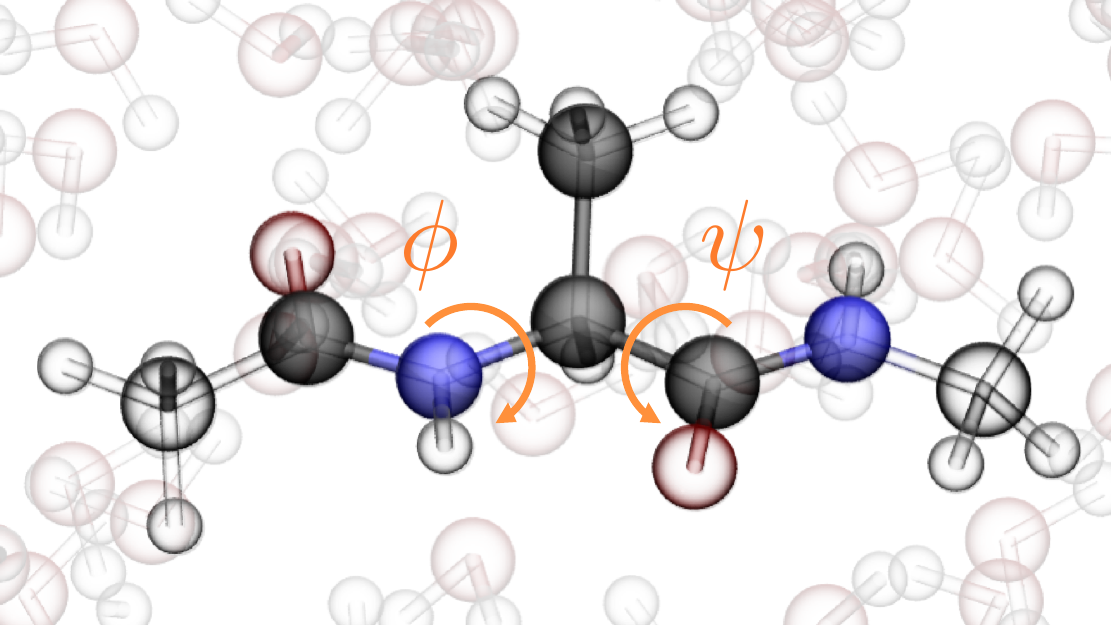}
\caption{
Capped alanine in water. The six shaded atoms are the ones preserved in the coarse-grained representation. The $\phi$ and $\psi$ dihedral angles completely describe the central alanine's heavy-atom dynamics.
}
\label{fig:ala2_struc}
\end{figure}

\begin{figure}[h!]
\centering
\includegraphics[width=0.5\textwidth]{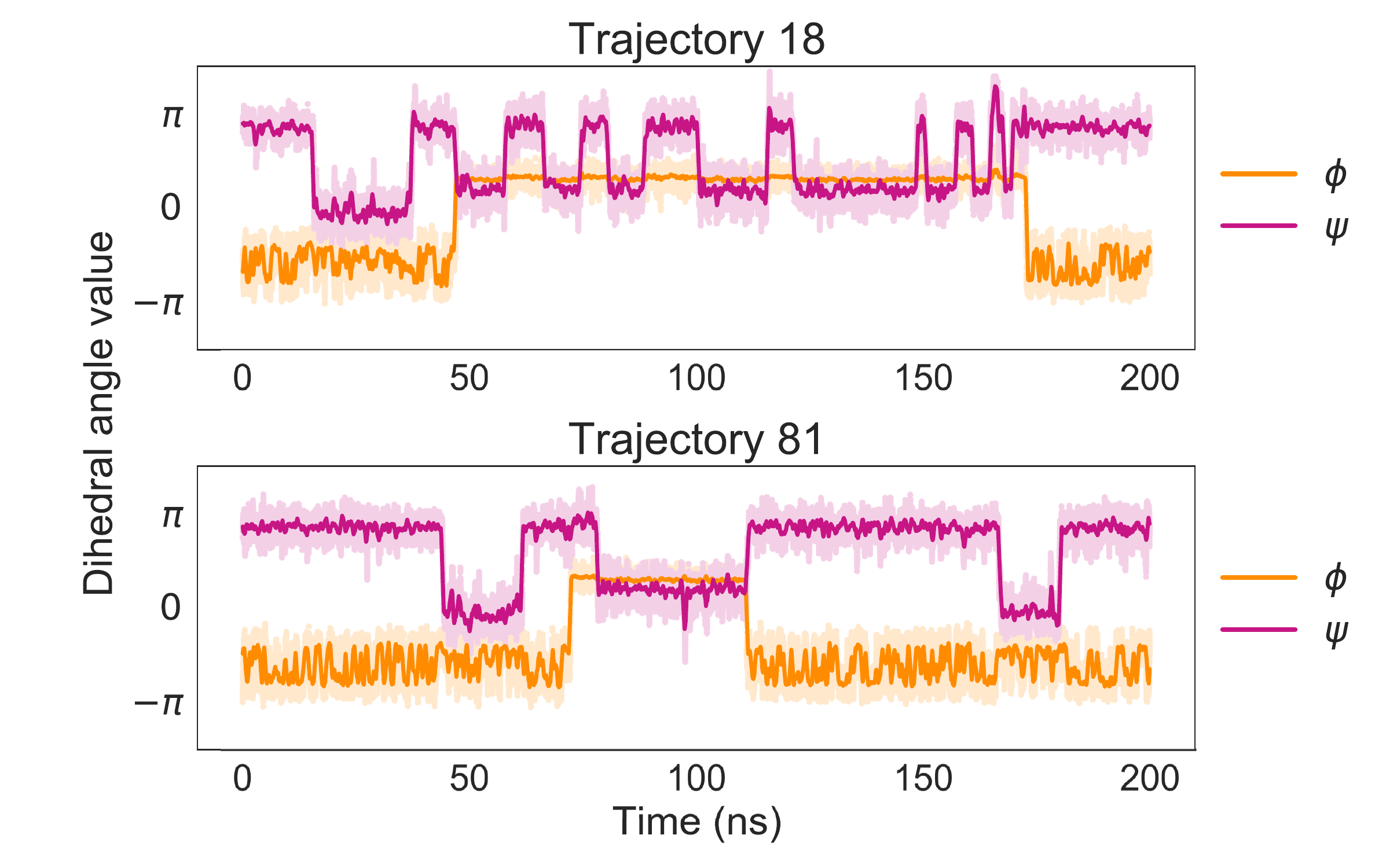}
\caption{
Two 100-ns trajectories simulated using a CGSchNet trained on atomistic data of capped alanine.
The orange and magenta lines represent the value of the dihedral angles $\phi$ and $\psi$, respectively, over the course of each simulation. 
Relatively steep changes in the $y$-direction indicate transitions among basins; one can see that both trajectories feature multiple transitions in both reaction coordinates.
A moving average of 250 simulation frames is used to smooth the darker curves.
}
\label{fig:ala2_indiv_ps}
\end{figure}

\begin{figure*}[t!]
\centering
\includegraphics[width=\textwidth]{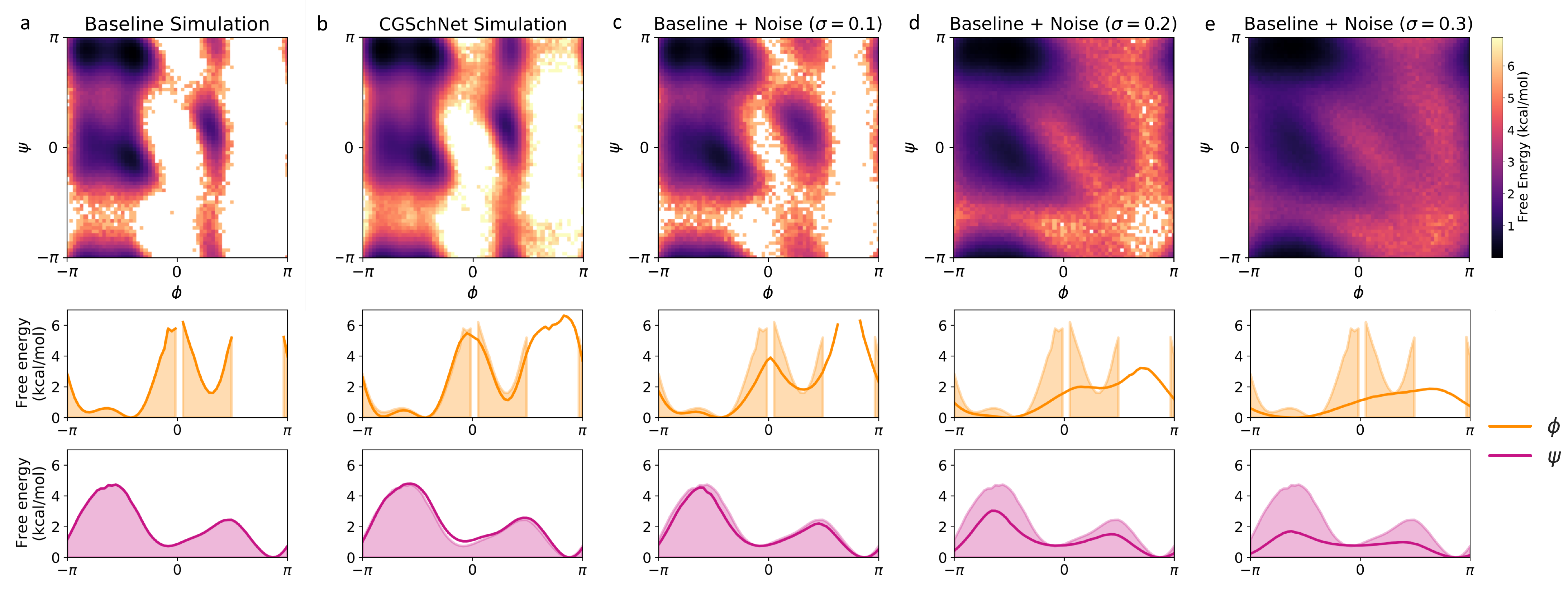}
\caption{
Two- and one-dimensional free energy surfaces for five capped alanine datasets.
From left to right, datasets are the baseline all-atom capped alanine simulation (a), the coarse-grained CGSchNet simulation produced for analysis (b), and datasets generated from perturbations of the original Cartesian coordinates of the baseline dataset drawn from noise distributed as $\mathcal{N}(0, \sigma^2)$ for $\sigma =$  0.1~\AA~(c), 0.2~\AA~(d), and 0.3~\AA~(e).
To create each two-dimensional surface, the $\phi$ and $\psi$ Ramachandran angles are calculated from the spatial coordinates and discretized into 60 $\times$ 60 regularly spaced square bins. The bin counts are converted to free energies by taking the natural log of the counts and multiplying by $-k_B T$; the color scale is the same in all five two-dimensional surfaces and darker color represents lower free energy (i.e., greater stability).
To obtain the one-dimensional $\phi$ and $\psi$ landscapes, free energies are calculated for $60$ regularly spaced bins along the reaction coordinate.
The shaded region always represents the baseline dataset and the bold line represents the dataset indicated in the subfigure title.
}
\label{fig:ala2_5x}
\end{figure*}

\begin{figure}[h!]
\centering
\includegraphics[width=0.5\textwidth]{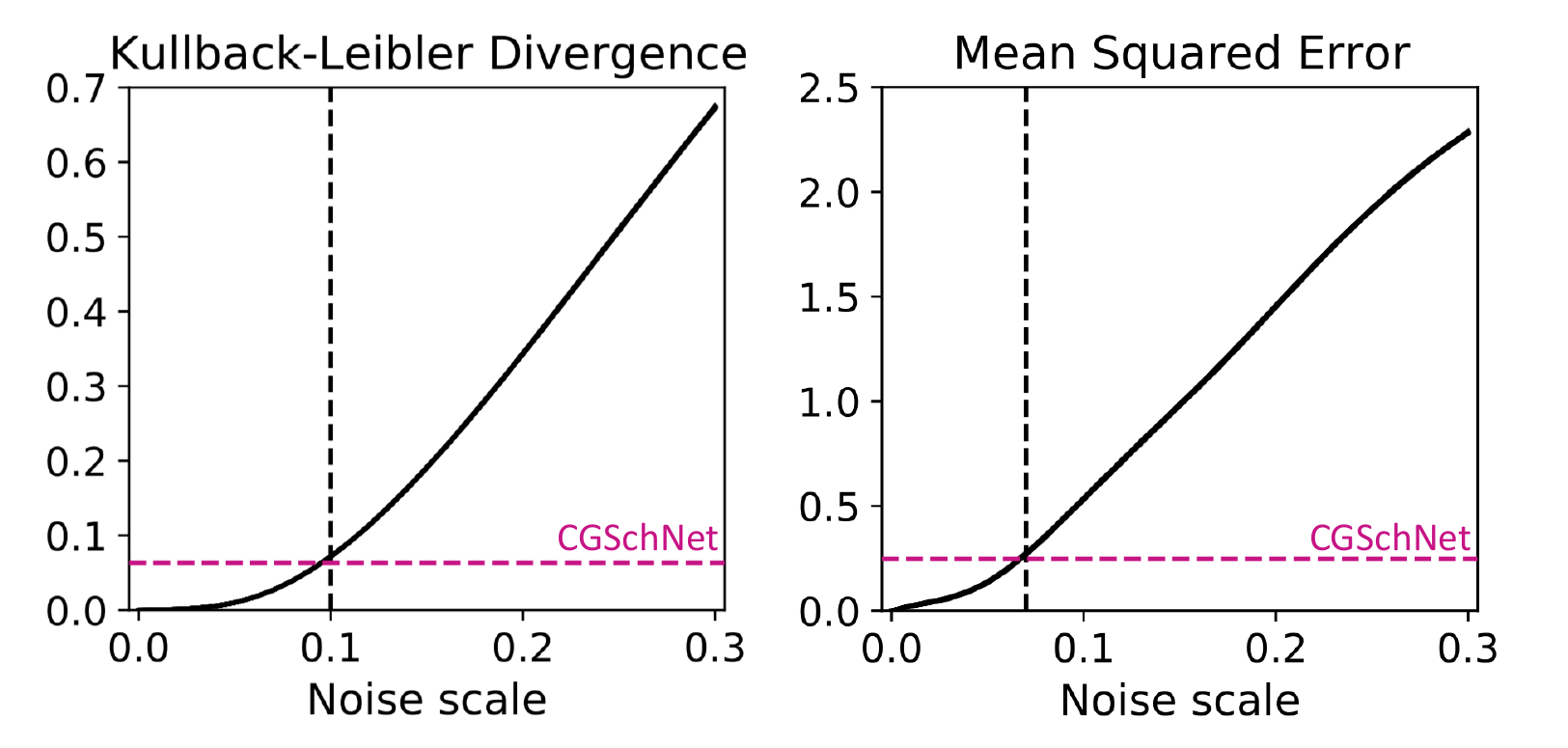}
\caption{
The Kullback-Leibler divergence (left) and mean squared error (right) are calculated between the baseline capped alanine dataset and the datasets obtained from perturbations to the baseline simulation at noise scale values of $\sigma \in \{0, 0.01, 0.02, \dots, 0.30\}$ where the former is the reference distribution and the latter is the trial distribution.
This procedure is performed 50 times with different random seeds; both plots show the superposition of those 50 lines.
The colored horizontal dashed line shows the value of the metric when comparing the CGSchNet simulation to the baseline and the black vertical dashed line indicates the noise scale $\sigma$ that return the closest value for that metric.
The KL divergence is computed for normalized bin counts in $\phi$ $\times$ $\psi$ space, and the MSE is computed for the energies of those bins as described in the main text.
}
\label{fig:ala2_div}
\end{figure}

\section{Results} \label{sec:results}

\subsection{Capped alanine} \label{subsec:alanine}

Capped alanine---often referred to alanine dipeptide for its two peptide bonds---is a common benchmark for MD methods development because the heavy-atom dynamics of the central alanine are completely described by the dihedral (torsional) angles $\phi$ and $\psi$ (see Fig.~\ref{fig:ala2_struc}).
We performed a single 1-$\mu$s all-atom, explicit solvent MD simulation for capped alanine and saved the forces to use for CGSchNet training (see Ref.~\onlinecite{wang2019machine} and supplementary material Sec.~A).
We can visualize the occupancies of backbone angle conformations by creating a histogram of the data on $\phi$ $\times$ $\psi$ space and visualizing the populations of the histogram bins.
This is called a Ramachandran map and is depicted in Fig.~\ref{fig:ala2_5x}a for the atomistic simulation using a $60 \times 60$ regular spatial discretization.

As an initial benchmark of the CGSchNet method, we aim to learn a force field for a coarse-grained representation of capped alanine such that we can reproduce its heavy-atom dynamics using a trained CGSchNet instead of a more expensive explicit solvent all-atom MD simulation.
For our coarse-grained mapping, we select the backbone heavy atoms C--[N--C$_\alpha$--C]$_\text{Ala}$--N as well as the alanine C$_\beta$ for a total of six beads.\footnote{
As in Ref.~\onlinecite{wang2020ensemble}, we require the $\beta$-carbon in order to break the symmetry of the system (i.e., to enforce chirality).
In their demonstration of CGnet, \citet{wang2019machine} used only the five backbone heavy atoms as beads because chirality is enforced through dihedral features, which we do not use here.
}
We use atomic numbers for the bead embeddings as in the original SchNet formulation~\cite{schutt2017schnet}.
A CGSchNet is trained on the coordinates and forces of the all-atom simulation depicted in Fig.~\ref{fig:ala2_5x}a.
The learning procedure involves a hyperparameter selection routine and the training of multiple models under five-fold cross-validation for each hyperparameter set (see supplementary material Sec.~B).

Once a final architecture has been selected, the trained model can serve as a force field in the coarse-grained space; i.e., by predicting the forces on a set of input coarse-grained coordinates.
Along with an integrator, predicted forces can be used to propagate coarse-grained coordinates forward in time (recall Sec.~\ref{subsec:simulation}).
This procedure (i.e., force prediction with CGSchNet followed by propagation with an integrator) is iterated until a simulation dataset of the desired duration has been obtained.
Since we employ five-fold cross-validation during the model training procedure, we have five trained CGSchNet models with a common architecture at hand.
To perform our coarse-grained simulation, we simultaneously predict the forces on each set of input coordinates from all five trained networks, and the mean force vector is used to propagate Langevin dynamics according to~\eqref{eq:langevin}.

To facilitate sampling, 100 coarse-grained simulations of length 200~ns each are performed in parallel from various starting positions in Ramachandran space (see supplementary material Sec.~C and Fig.~S3).
The time series of the $\phi$ and $\psi$ values for two of the trajectories that feature transitions among the major basins are plotted in Fig.~\ref{fig:ala2_indiv_ps}.
The same trajectories are also overlaid on the two-dimensional energy surface in Fig.~S4 in the supplementary material.

Free energy surfaces resulting from the coarse-grained simulation dataset are presented in Fig.~\ref{fig:ala2_5x}b.
We can see qualitatively that the two-dimensional free energy surface from the CGSchNet simulation captures the same basins as the surface calculated from the baseline all-atom simulation.
In the one-dimensional free energy surfaces, we see that the barriers are well-approximated by the CGSchNet simulation data.

To calibrate our understanding of the CGSchNet simulation dataset's relationship to the baseline atomistic simulation dataset, we create a set of new systems by perturbing the Cartesian coordinates of the latter with noise distributed as $\mathcal{N}(0, \sigma^2)$ for $\sigma \in \{0, 0.01, 0.02, \dots, 0.30\}$~\AA.
From the perturbed Cartesian coordinates, the new $\phi$ and $\psi$ dihedrals are calculated and assigned to the same $60 \times 60$ regularly spaced bins in Ramachandran space.
Examples of the perturbed free energy surfaces are shown in Fig.~\ref{fig:ala2_5x}c, d, and e for $\sigma = $ 0.1~\AA, 0.2~\AA, and 0.3~\AA, respectively.
We see that the surfaces become smeared and the free energy barriers are reduced with increasing noise.

This ensemble of perturbed simulation datasets enables us to understand the CGSchNet-produced simulation in the context of the baseline atomistic simulation.
To quantify the relationship between two distributions, we can use the Kullback-Leibler (KL) divergence~\cite{kullback1951information} and a mean squared error (MSE) formulation.
The KL divergence is defined for discrete distributions as,

\begin{align} \label{eq:kl}
    -\sum_{i}^m p_i \ln \frac{q_i}{p_i}, \forall p_i \geq 0,
\end{align}
    
\noindent{}where $p$ and $q$ are the ``reference'' and ``trial'' distributions, respectively, and $m$ is the number of bins in each discrete distribution.
In this case, $p$ and $q$ represent the normalized bin \emph{counts}.
The index $i$ returns the normalized count from the $i$th bin of a 60 $\times$ 60 regular discretization of $\phi \, \times \, \psi$ space.
The distribution obtained from the baseline atomistic simulation always serves as the reference. 
The mean squared error used here is,

\begin{align} \label{eq:mse}
    \frac{1}{m'}\sum_i^{m'} (P_i - Q_i)^2 \ni p_i q_i > 0,
\end{align}

\noindent{}where $p_i$ and $q_i$ remain the normalized bin \emph{counts} and $P_i$ and $Q_i$ represent, respectively, the corresponding discrete distributions of bin \emph{energies} calculated as, e.g., $P_i = k_BT \log p_i$ for Boltzmann's constant $k_B$ and absolute temperature $T$.
When no count is recorded for a bin in either $p_i$ or $q_i$, those bins are omitted from the mean.
$m'$ represents the number of bins in which $p_i q_i > 0$ (i.e., both have finite energies).\footnote{
We could alternatively compute the mean squared error between discrete distributions of counts without omitting any bins.
}
50 different trials are performed at different random seeds for the full set of noise scales (i.e., at each noise scale for a given trial, values are drawn from $\mathcal{N}(0, \sigma^2) \in \mathbb{R}^{M \times 3n}$, where $M$ is the length of the trajectory dataset and $n$ is the number of coarse-grained beads).
Within each trial, at each noise scale value $\sigma$, the KL divergence and MSE are calculated.
The results are presented in the left plot in Fig.~\ref{fig:ala2_div}.

We see in Fig.~\ref{fig:ala2_div} that as the noise increases, both divergence metrics also increase.
The dashed lines in Fig.~\ref{fig:ala2_div} show us that the error on the CGSchNet simulation dataset is approximately comparable to the corresponding error on the perturbed dataset with noise scale $\sigma = 0.1$~\AA~(Fig.~\ref{fig:ala2_5x}c)~\footnote{Similar results were obtained for the two-dimensional Wasserstein distance.}.
Upon qualitative comparison of the free energy surfaces, however, the former has more visual fidelity to the baseline surface in Fig.~\ref{fig:ala2_5x}a than to the broader spread seen (and expected) in the latter.
We know that coarse graining can result in increased population in transition regions that are rarely visited in an all-atom model; this is what we observe in Fig.~\ref{fig:ala2_5x}b.
As a corollary, we do \emph{not} expect coarse graining to result in the absence of states known to exist in the baseline system.

\begin{figure}[h!]
\centering
\includegraphics[width=0.35\textwidth]{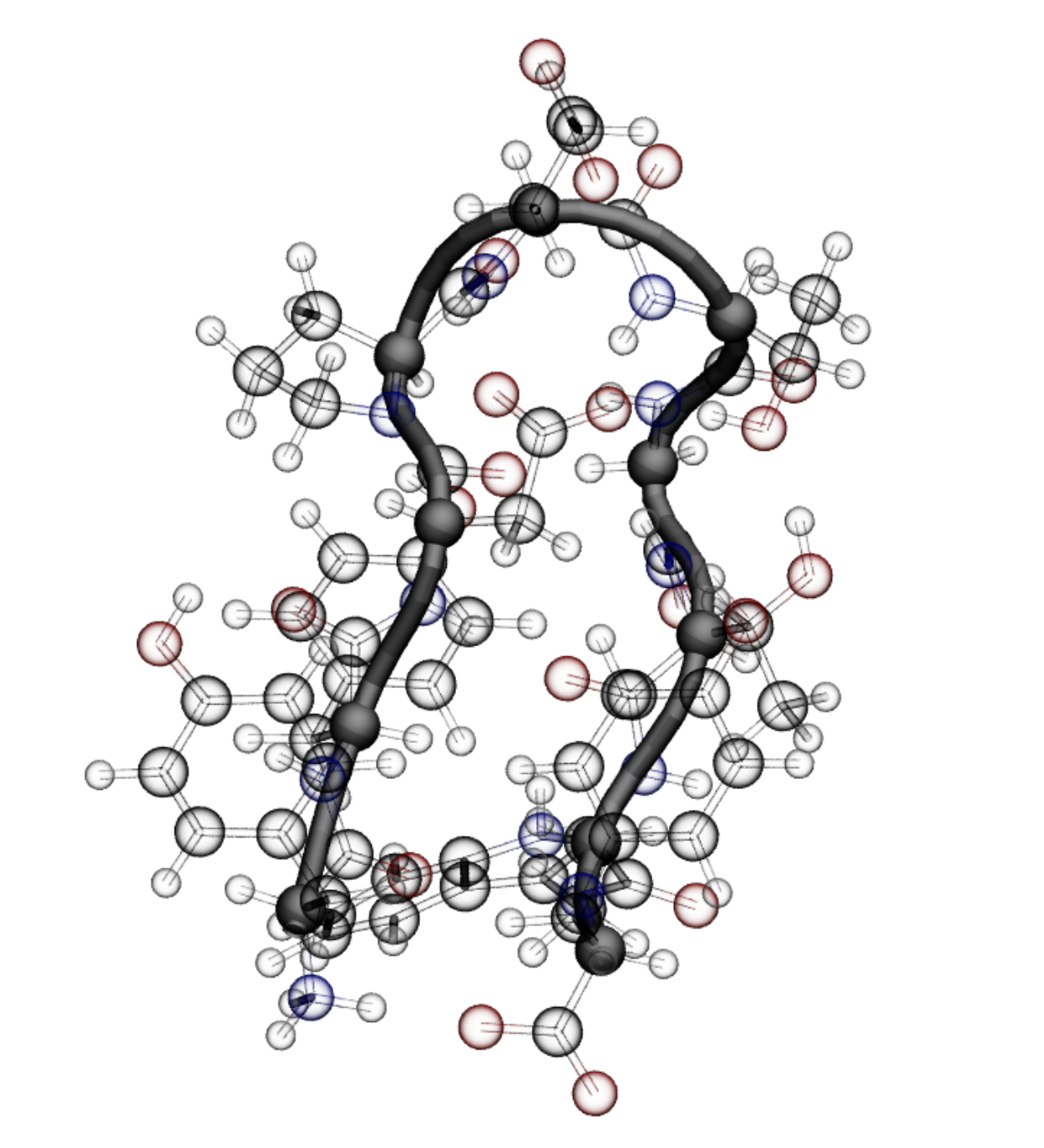}
\caption{
The miniprotein chignolin. The $\alpha$-carbon backbone is visualized in opaque black, and these ten atoms are the only beads preserved in the coarse-grained representation. The atomistic system is also solvated, although the water molecules are not shown here.
}
\label{fig:cln_struc}
\end{figure}

\subsection{Chignolin} \label{subsec:chignolin}

The CLN025 variant of chignolin is a 10-amino acid miniprotein~\cite{honda2008crystal} featuring a $\beta$-hairpin turn in its folded state (Fig.~\ref{fig:cln_struc}).
Due to its fast folding, its kinetics have been investigated in several MD studies~\cite{lindorff2011fast, beauchamp2012simple, husic2016optimized, mckiernan2017modeling, sultan2018automated, scherer2019variational}.
Our training data is obtained from an atomistic simulation of chignolin in explicit solvent for which we stored the forces (see Ref.~\onlinecite{wang2019machine} and supplementary material Sec.~A).
To build our CGSchNet model, we retain only the ten $\alpha$-carbons for our coarse-grained beads.
For the SchNet embeddings, we assign each amino acid type its own environment with a separate designation for the two terminal tyrosines.
After determining hyperparameters for our CGSchNet model, we simulate chignolin in the coarse-grained space using Langevin dynamics~\eqref{eq:langevin} as in the previous section.
The procedures for CGSchNet training and simulation are similar to those used for capped alanine and are described in the supplementary material Secs.~B and~C.

Given our CGSchNet simulation data, we are interested not only in performing a similar analysis to the one in the previous section for alanine dipeptide (i.e., comparison to the baseline dataset with and without noise added) but also to simulation data obtained from a CGnet trained according to the protocol in Ref.~\onlinecite{wang2019machine} for the same system.
We thus also construct a CGnet according to the parameters selected in Ref.~\onlinecite{wang2019machine} (i.e., using fixed geometric features as described in Sec.~\ref{subsec:cgnet_architecture}; see also Sec.~B in the supplementary material).
Then, we create a simulation dataset using the protocol described in the previous section and supplementary material Sec.~C.
Finally, we employ a similar protocol to the previous section by perturbing the raw Cartesian coordinates of the all-atom chignolin simulation dataset with noise distributed as $\mathcal{N}(0, \sigma^2)$ for $\sigma \in \{0, 0.03, 0.06, \dots, 0.90 \}$~\AA. 

For each type of system (i.e., baseline, CGSchNet, CGnet, and baseline with noise perturbation), we build Markov state models (MSMs)~\cite{zwanzig1983classical, schutte1999direct, swope2004describing, singhal2004using, chodera2007automatic, noe2007hierarchical, buchete2008coarse, prinz2011markov} (see Refs.~\onlinecite{husic2018markov} and ~\onlinecite{noe2020chapter} for recent overviews).
First, the data is ``featurized'' from Cartesian coordinates into the set of 45 distances between pairs of $\alpha$-carbons.
From these distances, time-lagged independent component analysis (TICA)~\cite{perez2013identification, schwantes2013improvements} is performed to yield four slow reaction coordinates for the system.
These four reaction coordinates are clustered into 150 discrete, disjoint states using the $k$-means algorithm.
An MSM is then estimated from the cluster assignments.
The MSM for the baseline simulation dataset is constructed first; then, the other simulation datasets are projected onto the space defined by the former.
MSM essentials are presented in supplementary material Sec.~D from a theoretical standpoint, and the specific protocols used for the MSM analysis in this section are given in supplementary material Sec.~E.

\begin{figure*}[t!]
\centering
\includegraphics[width=0.9\textwidth]{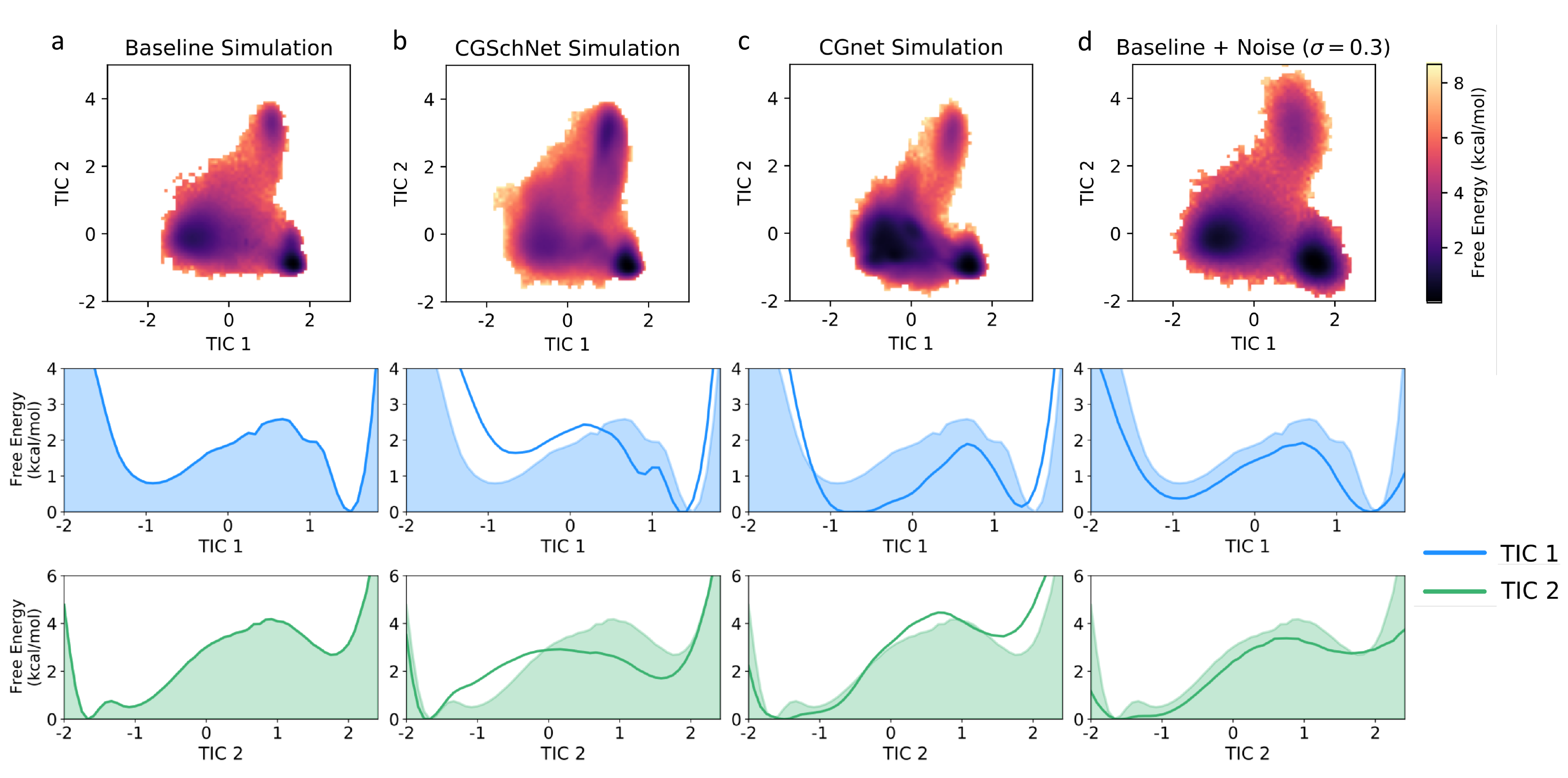}
\caption{
Two- and one-dimensional free energy surfaces for four chignolin datasets.
From left to right, datasets are the baseline chignolin simulation (a), the coarse-grained CGSchNet simulation (b), the coarse-grained CGnet simulation (c), and the dataset generated from the perturbation of the original Cartesian coordinates of the baseline dataset drawn from noise distributed as $\mathcal{N}(0, \sigma^2)$ for $\sigma =$  0.3~\AA~(d).
Each two-dimensional surface is obtained from a $120 \times 120$ histogram on TIC 1 $\times$ TIC 2 space with weights determined from the MSM built for each system (see main text).
The one-dimensional surfaces are similarly obtained from 120-bin histograms on a single TIC.
For each reweighted histogram bin, the free energy is obtained by taking the natural log of the counts and multiplying by $-k_B T$; the color scale is the same in all four two-dimensional surfaces and darker color represents lower free energy (i.e., greater stability).
The shaded region always represents the baseline dataset and the bold line represents the dataset indicated in the subfigure title.
}
\label{fig:cln_fes}
\end{figure*}

The stationary distribution of each MSM is then used to reweight the TICA coordinates used for its own construction.
Histograms of the first two TICA coordinates are presented in the top row of Fig.~\ref{fig:cln_fes} for the baseline, CGSchNet, and CGnet simulation datasets as well as the baseline dataset for $\sigma = 0.3$.
The first two reweighted TICA coordinates are also individually binned into one-dimensional free energy surfaces, which are depicted in the second and third rows of Fig.~\ref{fig:cln_fes}.
We see that the free energy barriers along these reaction coordinates are reasonably approximated by the CGSchNet simulation.

\begin{figure}[h!]
\centering
\includegraphics[width=0.45\textwidth]{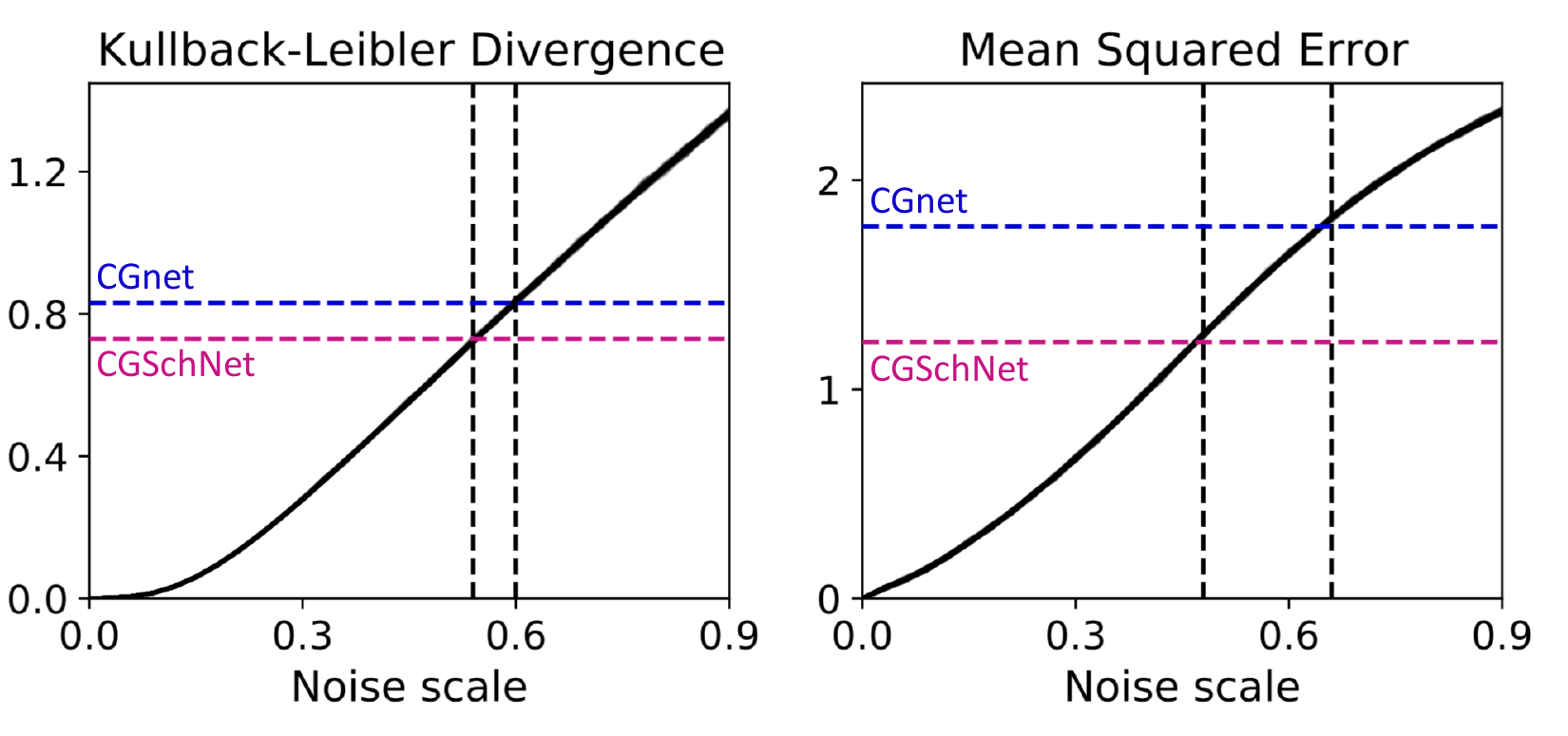}
\caption{
The Kullback-Leibler divergence (left) and mean squared error (right) are calculated between the baseline chignolin dataset and the datasets obtained from perturbations to the baseline simulation at noise scale values of $\sigma \in \{0, 0.03, 0.06, \dots, 0.90\}$ where the former is the reference distribution and the latter is the trial distribution.
This procedure is performed 50 times with different random seeds; both plots show the superposition of those 50 lines.
The colored horizontal dashed lines show the values of the metric when comparing the CGSchNet (purple) and the CGnet (blue) simulations to the baseline.
The black vertical dashed line indicates the noise scale $\sigma$ that return the closest value for that metric.
The KL divergence is computed for normalized bin counts in reweighted TIC 1 $\times$ TIC 2 space, and the MSE is computed for energies as described in the main text.
}
\label{fig:cln_div}
\end{figure}

Figure~\ref{fig:cln_div} shows the same divergence metrics calculated in Fig.~\ref{fig:ala2_div} in the previous section.
Again, we see that both the KL divergence and the MSE increase monotonically with the magnitude of the noise.
In this case, we can assess the equivalent noise value for both the CGSchNet and CGnet simulation datasets.
For both divergences measured, we see that the CGSchNet simulation corresponds to a lesser value of added noise than the CGnet simulation.

\begin{figure}[h!]
\centering
\includegraphics[width=0.45\textwidth]{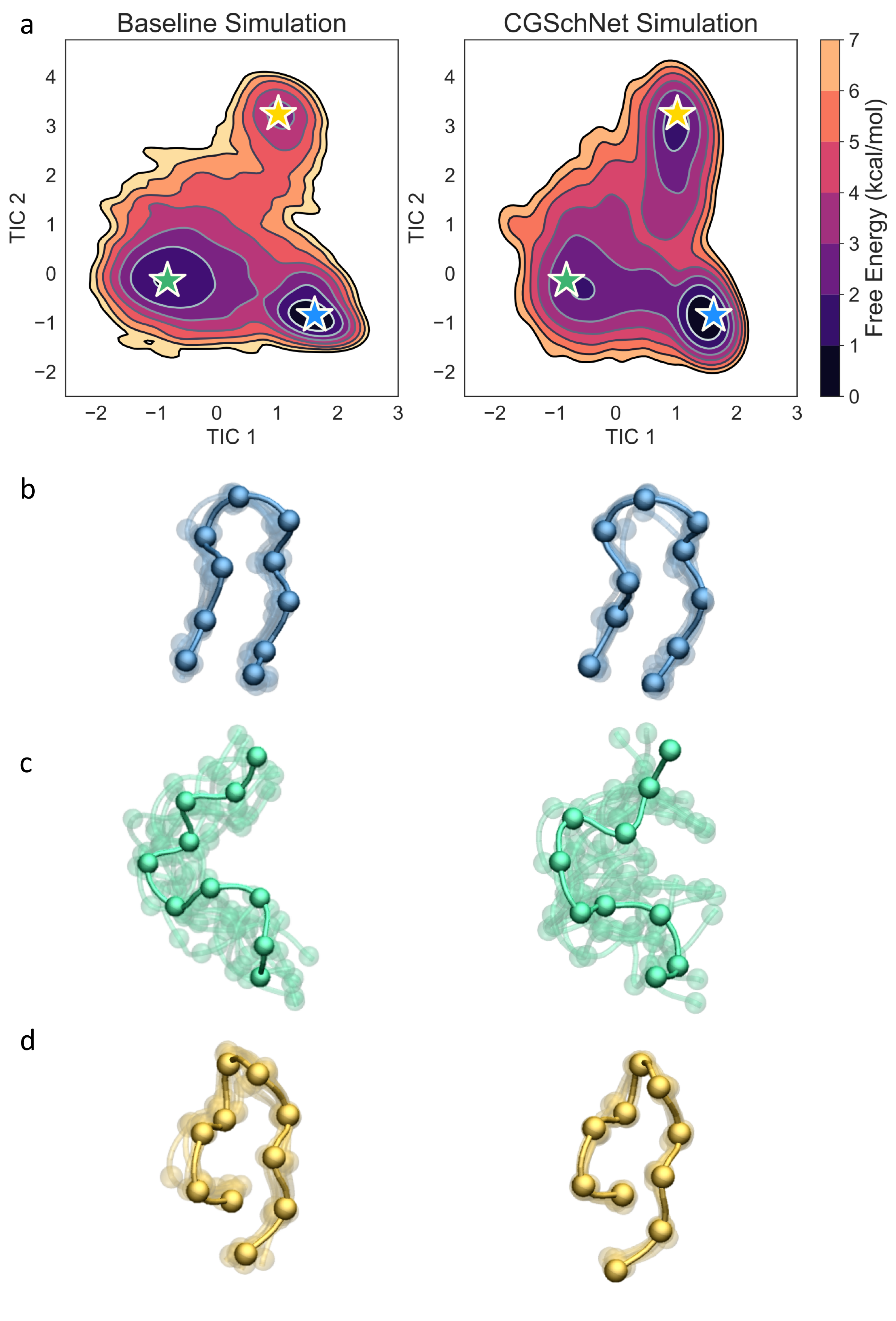}
\caption{
Two-dimensional free energy surfaces (a) and sample folded (b), unfolded (c), and misfolded (d) conformations from the baseline atomistic simulation of chignolin (left column) and the CGSchNet simulation (right column).
The free energy surfaces are built from 150-state MSMs that were constructed from the slowest two TICA coordinates in contact distance space.
The color scale is the same for both surfaces and darker color represents lower free energy (i.e., greater stability).
Each set of ten sampled structures corresponds to the MSM state represented by the star on the free energy surface of the same color (one of the ten structures is opaque for clarity).
}
\label{fig:cln_3stars}
\end{figure}

We can also obtain free energy surfaces from the MSMs constructed for the systems; the surfaces for the baseline and CGSchNet simulation datasets of chignolin are presented in Fig.~\ref{fig:cln_3stars}a on the left and right, respectively.
We see that the three major basins observed in the atomistic data are captured by CGSchNet.
These basins represent folded, unfolded, and misfolded ensembles and are indicated in Fig.~\ref{fig:cln_3stars}a with blue, green, and yellow stars, respectively.
Each star represents one of the 150 MSM states and was manually selected from the MSM states near the relevant basin (see the supplementary material Fig.~S8 for a visualization of all 150 MSM states).
To verify that the protein conformations are similar in each of the states, we sample ten structures from each starred state per simulation dataset.
The structures are visualized in Fig.~\ref{fig:cln_3stars}b-d, and the similarity of the structures on the left-hand side (baseline simulation) to those on the right-hand side (CGSchNet simulation) from corresponding MSM states is apparent.

The analysis of the CGSchNet and CGnet simulation datasets so far used TICA reaction coordinates that were obtained by projecting the simulation data onto coordinates defined by a TICA model built for the baseline atomistic data (see supplementary material Sec.~E).
This was done in order to compare simulation results using the same reaction coordinates.
We can also construct TICA models from the simulation data without projection to determine the scaling factor for the coarse-grained timescale.
For this analysis, we build two further (independent) TICA models for the CGSchNet and CGnet datasets at a lag time long enough for the TICA timescales to have leveled off (see Fig.~S9 in the supplementary material).
A 100-round bootstrapping analysis of the longest TICA timescale from the CGSchNet simulation data yields a time scaling factor of 2.2 with a standard deviation of 0.4.
From this time rescaling we determine that the effective collision rate (i.e., friction) of the coarse-grained simulations is 180--260~ps$^{-1}$.
This value is four orders of magnitudes larger than the friction constant in the all-atom model (0.1~ps$^{-1}$)~\cite{wang2019machine}, which we expect because we have coarse-grained out the solvent dynamics.
The same analysis for the CGnet simulation data yields a scaling factor of $2.2 \pm 0.3$ and a corresponding effective collision rate of 190--250~ps$^{-1}$.

Although we use MSMs and TICA models to obtain thermodynamics and effective friction constants, we do not attempt a kinetic analysis in the present work because the scope of force matching is limited to thermodynamic consistency~\cite{izvekov2005multiscale, noid2008multiscale}.
The matching of dynamics in addition to thermodynamics is an open challenge that has been the subject of recent work~\cite{nuske2019coarse}.
Given coarse-grained dynamics, analytical methods have been derived that enable their rescaling to the dynamics of the system's all-atom counterpart~\cite{lyubimov2011first}.

\section{Discussion}

Coarse graining holds the promise of simulating larger systems at longer timescales than are currently possible at the atomistic level.
However, mathematical frameworks must be developed in order to ensure that the results obtained from a coarse-grained model are faithful to those that would be obtained from an atomistic simulation or experimental measurement.
Force matching~\cite{ercolessi1994interatomic, izvekov2005multiscale} is one such framework that, when certain restrictions are applied, guarantees thermodynamic consistency with atomistic data in the variational limit~\cite{noid2008multiscale}.
Such a variational framework enables the formulation of the force matching problem as a supervised machine learning task, which is presented in Ref.~\onlinecite{wang2019machine} as CGnet.

A key limitation of the original CGnet is that it is not transferable across different systems:~a new network must be trained for each individual molecular system under study because the molecular features from which it learns the force field must be chosen by hand.
Here, we replace manually determined features with a learnable representation.
This representation is enabled by the use of continuous filter convolutions on a graph neutral network (i.e., SchNet~\cite{schutt2017schnet, schutt2018schnet}).
SchNet is an inherently transferable architecture originally designed to match energies and forces to quantum calculations for small organic molecules.
By leveraging SchNet in the \emph{coarse graining} context---i.e., to learn the molecular features input into a CGnet---we render the hybrid CGnet architecture (i.e., CGSchNet) transferable across molecular systems of different sizes and sequences.

Our aim in the present contribution is threefold:~to summarize the variational framework enabling a supervised learning approach to force matching, to provide an accompanying software package implementing the methods discussed herein (see Appendix~\ref{app:software}), and to demonstrate that CGSchNet produces results on individual systems that are superior to those obtained from bespoke features. 
The advances presented in this work prepare us to address the ultimate challenge of machine learning a coarse-grained force field that is transferable across molecular systems.

In our computational experiments performed on capped alanine and the miniprotein chignolin, we find that CGSchNet's performance exceeds that of CGnet in three ways.
First, the free energy surface obtained from CGSchNet simulations of chignolin is more accurate than the free energy surface presented for the same systems in Ref.~\onlinecite{wang2019machine}.
Second, CGSchNet is more robust to network hyperparameters than its predecessor.
In fact, for the CGSchNet hyperparameters varied during model training (see supplementary material Sec.~B), the same selections are used for both systems presented in Sec.~\ref{sec:results}.
Third, CGSchNet employs less regularization; particularly, it does not require the extra step of enforcing a Lipschitz constraint~\cite{gouk2018regularisation} on its network's weight matrices as was found to be necessary for CGnet~\cite{wang2019machine}.

While our current protocol has demonstrated success for a capped monopeptide and a 10-amino acid miniprotein, adapting the CGSchNet pipeline to produce accurate coarse-grained force fields for larger protein systems remains an open challenge.
Addressing this challenge may require specific sampling strategies when obtaining training data, the incorporation of new priors that inform tertiary structure formation, or modifications to the CGSchNet architecture itself such as regularization.
Successfully modeling the thermodynamics of protein folding or conformational change via a transferable, machine-learned force field would signify a major success for the union of artificial intelligence and the computational molecular sciences.

The method introduced herein enables us to reproduce the thermodynamics of small protein systems using an architecture that is transferable across system size and sequence.
However, CGSchNet is not readily transferable across thermodynamic states.
Related work leveraging the same variational principle in a semi-supervised learning context allows the learning of coarse-grained representations over multiple thermodynamic states, enabling transferability across different temperatures~\cite{wang2019coarse, ruza2020temperature}.
This method has been demonstrated for ionic liquids, for which nonequilibrium transport properties are of prime interest.
Finally, the reproduction of kinetics is an open research problem, and methods for the so-called ``spectral matching'' problem have recently been introduced~\cite{nuske2019coarse}.
Ideally, both force and spectral matching could be pursued in conjunction to match both thermodynamics and kinetics simultaneously.

Structural, bottom up coarse graining consists of two aspects:~the model resolution and the force field.
Here, we assume the resolution is set and focus on the force field, but the choice of an optimal model resolution is itself a significant challenge that is interconnected to the goal of force field optimization.
How to choose a resolution for coarse graining---and the interplay of this choice with transferable force field architectures---remains an open question.
Recent work has employed machine learning and data-driven approaches to pursue an optimal resolution using various objectives~\cite{boninsegna2018data, wang2019coarse}.

Altogether, the methodology we introduce in the present contribution establishes a transferable architecture for the machine learning of coarse-grained force fields, and we expect our accompanying software to facilitate progress not only in that realm but also towards the outstanding challenges of learning coarse-grained dynamics and optimizing a model's resolution.


\section*{Supplementary Material}

See the supplementary material for further specifics on simulation, model training, and MSM construction.

\section*{Acknowledgements}

B.E.H.~is immeasurably grateful to Moritz Hoffmann for his wisdom and support.
The authors are grateful to Dr.~Ankit Patel for discussions on model debugging and machine learning techniques, to Iryna Zaporozhets for discussions concerning model priors, to Dr.~Stefan Doerr for software help, to Dr.~Jan Hermann for discussions regarding SchNet, and to the reviewers for their constructive comments.

We acknowledge funding from the European Commission (ERC CoG 772230 ``ScaleCell'') to B.E.H., A.K, Y.C., and F.N;
from MATH+ the Berlin Mathematics Center to B.E.H.~(EF1-2), S.O.~(AA1-6), and F.N (EF1-2 and AA1-6);
from the Natural Science Foundation (CHE-1738990, CHE-1900374, and PHY-1427654) to N.E.C., J.W., and C.C.;
from the Welch Foundation (C-1570) to N.E.C, J.W., and C.C.;
from the NLM Training Program in Biomedical Informatics and Data Science (5T15LM007093-27) to N.E.C.; 
from the Einstein Foundation Berlin to C.C.;
and
from the Deutsche Forschungsgemeinschaft (SFB1114 projects A04 and C03) to F.N.
D.L.~was supported by a FPI fellowship from the Spanish Ministry of Science and Innovation (MICINN, PRE2018-085169).
G.D.F.~acknowledges support from MINECO (Unidad de Excelencia Mari{\'a} de Maeztu AEI [CEX2018-000782-M] and BIO2017-82628-P) and FEDER.
This project has received funding from the European Union’s Horizon 2020 research and innovation programme under Grant Agreement 823712 (CompBioMed2 Project).
We thank the GPUGRID donors for their compute time.

Simulations were performed on the computer clusters of the Center for Research Computing at Rice University, supported in part by the Big-Data Private-Cloud Research Cyberinfrastructure MRI-award (NSF grant CNS-1338099), and on the clusters of the Department of Mathematics and Computer Science at Freie Universit{\"a}t, Berlin. 

Part of this research was performed while B.E.H., N.E.C., D.L., J.W., G.dF., C.C., and F.N.~were visiting the Institute for Pure and Applied Mathematics (IPAM) at the University of California, Los Angeles for the Long Program ``Machine Learning for Physics and the Physics of Learning.''
IPAM is supported by the National Science Foundation (Grant No.~DMS-1440415).

\section*{Data Availability Statement}
The data that support the findings of this study are available from the corresponding authors upon reasonable request.

\appendix


\section{Software} \label{app:software}

The \texttt{cgnet} software package is available at \url{https://github.com/coarse-graining/cgnet} under the BSD-3-Clause license.
\texttt{cgnet} requires NumPy~\cite{numpy}, SciPy~\cite{scipy}, and PyTorch~\cite{paszke2019pytorch}, and optional functionalities further depend on pandas~\cite{pandas}, MDTraj \cite{mcgibbon2015mdtraj}, and Scikit-learn~\cite{scikit-learn}.
The examples are provided in Jupyter notebooks~\cite{jupyter} which also require Matplotlib~\cite{matplotlib}.
The SchNet part of the code is inspired by SchNetPack~\cite{schutt2018schnetpack} and the Langevin dynamics simulation code is adapted from OpenMM~\cite{eastman2017openmm}.
In addition to \texttt{cgnet} and the packages already mentioned, visualization was aided by Seaborn~\cite{seaborn} and VMD~\cite{vmd}.
Analysis was facilitated by PyEMMA~\cite{scherer2015pyemma, wehmeyer2018introduction}.

\bibliography{refs}

\end{document}